\documentclass[11pt,a4paper]{article}
\usepackage{jcappub}

\usepackage{appendix}
\usepackage{array}
\newcolumntype{x}[1]{>{\centering\arraybackslash}p{#1}}

\ifx\pdfoutput\undefined
\usepackage[dvips,bookmarks=false]{hyperref}	% This is for arXiv.org
\else
\usepackage{hyperref}	% This is for pdftex
\fi
\hypersetup{colorlinks,bookmarksopen,bookmarksnumbered,citecolor=blue,
linkcolor=black,pdfstartview=FitH,urlcolor=blue}

\usepackage{graphicx}
\usepackage{amsmath}
\usepackage{amssymb}

\newcommand{\be}{\begin{equation}}
\newcommand{\ee}{\end{equation}}
\newcommand{\beq}{\begin{equation}}
\newcommand{\eeq}{\end{equation}}

\newcommand{\vect}[1]{\boldsymbol{\rm #1}}
\title{Inverted dipole feature in directional detection of exothermic dark matter}

\author[a]{Nassim Bozorgnia,}
\author[b]{Graciela B. Gelmini}
\author[c]{and Paolo Gondolo}
\affiliation[a]{GRAPPA, University of Amsterdam,\\
Science Park 904,
  1090 GL Amsterdam, Netherlands} 
\affiliation[b]{Department of Physics and Astronomy, UCLA,\\
475 Portola Plaza, Los Angeles, CA 90095, USA}
\affiliation[v]{Department of Physics and Astronomy, University of Utah,\\
115 South 1400 East \#201, Salt Lake City, UT 84112, USA}
\emailAdd{n.bozorgnia@uva.nl}
\emailAdd{gelmini@physics.ucla.edu}
\emailAdd{paolo@physics.utah.edu}

\abstract{
Directional dark matter detection attempts to measure the direction of motion of nuclei recoiling after having interacted with dark matter particles in the halo of our Galaxy. Due to Earth's motion with respect to the Galaxy, the dark matter flux is concentrated around a preferential direction. An anisotropy in the recoil direction rate is expected as an unmistakable signature of dark matter. The average nuclear recoil direction is expected to coincide with the average direction of dark matter particles arriving to Earth. Here we point out that for a particular type of dark matter, inelastic exothermic dark matter,  the mean recoil direction as well as a secondary feature, a ring of  maximum recoil rate around the  mean recoil direction, could instead be opposite to the average dark matter  arrival direction. Thus, the detection of an average nuclear recoil direction opposite to the usually expected direction would constitute a spectacular experimental confirmation of this type of dark matter. 
 }

\arxivnumber{}

\begin{document}
\maketitle

\section{Introduction}
\label{sec:intro}

Almost 85\% of the matter of our own Galaxy is in a dark matter (DM) halo which extends much beyond the visible disk of the Galaxy. Direct DM experiments search for energy deposited in the scattering off nuclei in low-background detectors  of Weakly Interacting Massive Particles (WIMPs) present in the dark halo of our Galaxy. Some direct DM detectors attempt  to measure also the direction of the  recoiling nucleus besides its energy~\cite{Mayet:2016zxu}. Because  WIMPs arrive to us from a preferential direction, an anisotropy in the recoil direction rate is expected which would constitute  an unmistakable signature  of DM. The easiest DM signature to detect in directional detectors is the average recoil direction. Here we point out that for a particular type of DM, inelastic exothermic DM,  the mean recoil direction as well as a secondary feature, a ring of  maximum recoil rate around the  mean recoil direction~\cite{Bozorgnia:2011vc}, could be opposite to the average arrival direction of WIMPs at any particular time. Thus the detection of a  nuclear recoil direction opposite to the usually expected direction would constitute a spectacular experimental confirmation of this type of DM. 

 For inelastic DM~\cite{TuckerSmith:2001hy}, the incoming DM particle $\chi$ scatters dominantly to a different DM state  $\chi^*$ with a mass difference $\delta = m_{\chi^*}- m_\chi$,  where $|\delta| < < m$, and $\delta > 0$ and $<0$ respectively describe endothermic~\cite{TuckerSmith:2001hy}  and exothermic~\cite{Finkbeiner:2007kk,Batell:2009vb,Graham:2010ca} scattering.

 In this type of models part of the DM consists of the $\chi$ state and part of it of the $\chi^*$. In the exothermic scattering the DM down-scatters in mass, but in principle also the endothermic scattering, in which the DM up-scatters in mass, is possible.  For the inelastic endothermic scattering of an incoming $\chi^*$ into a final particle $\chi$ not to contribute to the signal, it must be kinematically forbidden because the required incoming  speed would be larger than the escape speed from our Galaxy, $v_{\rm esc}$. $v_\delta$ is the minimum  possible value of the minimum speed $v_{\rm min} $ required for the WIMP to impart a recoil energy $E_R$ to the nucleus.  For $\delta < 0$, $v_\delta$ =0. For $\delta > 0$,  $v_\delta = \sqrt{2 \delta / \mu}$, where $\mu$ is the reduced mass of the  DM-nucleus system. Thus only if $v_\delta = \sqrt{2 \delta / \mu}> v_{\rm esc} $ the  scattering is purely exothermic.
 
The paper is structured as follows. In section~\ref{sec:notation} which is the main section of this paper, we present the kinematics of inelastic DM scattering and illustrate the effect of the inverted dipole feature for exothermic scattering, and in section~\ref{Conclusions} we conclude. In appendix~\ref{app-A} we present the derivation of the Radon transform for inelastic scattering.

\section{Kinematics of inelastic dark matter scattering}
\label{sec:notation}

 For inelastic DM, $v_q$, the projection of the  incoming lab frame WIMP velocity $\vect{v}$  in the direction of the recoil momentum ${\hat{\bf q}}$  is given by,
\beq
v_q= \vect{v} \cdot \hat{\vect q}= \sqrt{\frac{1}{2 m_T E_R}}\left( \frac{m_T E_R}{\mu} + \delta \right),
\label{eq:vq-inelastic}
\eeq
where $m_T$ is the mass of the target nucleus. 
Notice that $v_q$ is negative for negative values of $\delta < - ({m_T E_R}/{\mu})$, which means that the recoil momentum
$\vect{v}$ forms an angle larger than 90 degrees with the incoming WIMP velocity, i.e. the nucleus recoils backwards.  

 The  minimum speed the DM particle must have  to deposit a recoil energy $E_R$ in the detector is always
\beq
v_{\rm min}= \left| v_q  \right|.
\label{eq:vmin-inelastic}
\eeq

The distribution of nuclear recoil directions $dR/dE_Rd\Omega_{q}$ is proportional to the Radon transform of the WIMP velocity distribution (neglecting the nuclear form factor dependence on $E_R$). The form of the Radon transform~\cite{Gondolo:2002np} of the DM velocity distribution $f({\bf v})$ for inelastic scattering is the same as for elastic scattering~\cite{Finkbeiner:2009ug, Barello:2014uda} (see also appendix~\ref{app-A}),
\beq
\hat{f}(v_q, \hat{\vect v}_q) = \int \delta({\vect v} \cdot \hat{\vect{v}}_q- v_q) f({\bf v}) d^3 v.
\label{eq:Radon}
\eeq
Notice that this equation is many times written in the literature with $v_{\rm min}$ in place of $v_q$, and this is correct except when 
$v_q$ is negative, which can only happen in exothermic scattering. The Radon transform in the laboratory frame for the Maxwellian WIMP velocity distribution truncated at the escape speed $v_{\rm esc}$ (with respect to the Galaxy) is given by~\cite{Gondolo:2002np}
\begin{equation}
\label{eq:fhatTM}
\hat{f}\!\left( v_q, \hat{\bf q} \right)=\frac{1}{{N_{\rm esc}(2\pi \sigma_v^2)^{1/2}}}~{\left\{\exp{\left[-\frac{\left[ v_q + \hat{\bf q} \cdot {\bf V}_{\rm lab}\right]^2}{2\sigma_v^2}\right]}-\exp{\left[-\frac{v_{\rm esc}^2}{2\sigma_v^2}\right]}\right\}},
\end{equation}
if $v_q + \hat{\bf q} \cdot {\bf V}_{\rm lab} < v_{\rm esc}$, and zero otherwise. Here $\hat{\bf q}$ is the unit vector in the direction of the nuclear recoil momentum ${\bf q}$, ${\bf V}_{\rm lab}$ is the velocity of the laboratory with respect to the Galaxy (hence the average velocity of the WIMPs with respect to the detector is $-{\bf V}_{\rm lab}$), $\sigma_v$ is the DM velocity dispersion, and
\begin{equation}
N_{\rm esc} = {\rm erf}{\left(\frac{v_{\rm esc}}{\sqrt{2}\sigma_v} \right)} - \sqrt{\frac{2}{\pi}}\frac{v_{\rm esc}}{\sigma_v} \exp{\left[-\frac{v_{\rm esc}^2}{2 \sigma_v^2} \right]}.
\end{equation}

The maximum of $\hat{f}$ in eq.~\eqref{eq:fhatTM} occurs at an angle $\gamma$ between $\hat{\bf q}$  and  the average velocity of the incoming WIMPs, $-{\bf V}_{\rm lab}$, given by
\beq
\gamma = 
\begin{cases}
\pi  & \text{for } v_q \le -V_{\rm lab} , \\
\arccos\left( \displaystyle \frac{v_q}{V_{\rm lab}} \right)  & \text{for } - V_{\rm lab} < v_q < V_{\rm lab}, \\
0  & \text{for } v_q \ge V_{\rm lab}.
\end{cases}
\eeq
Thus for $- V_{\rm lab} < v_q < V_{\rm lab}$, there is a ring of angular radius $\gamma$ on which the number of recoil events is maximal, with the ring center in the direction of $-{\bf V}_{\rm lab}$. Hence, for a given recoil energy $E_R$ the ring angular radius $\gamma$ is given by
\beq
\cos\gamma =  \frac{v_q}{V_{\rm lab}} = \sqrt{\frac{1}{2 m_T E_R V_{\rm lab}^2}}\left[ \frac{m_T E_R}{\mu} + \delta \right].
\label{gamma}
\eeq

\begin{figure}
\centering\includegraphics[width=0.49\textwidth]{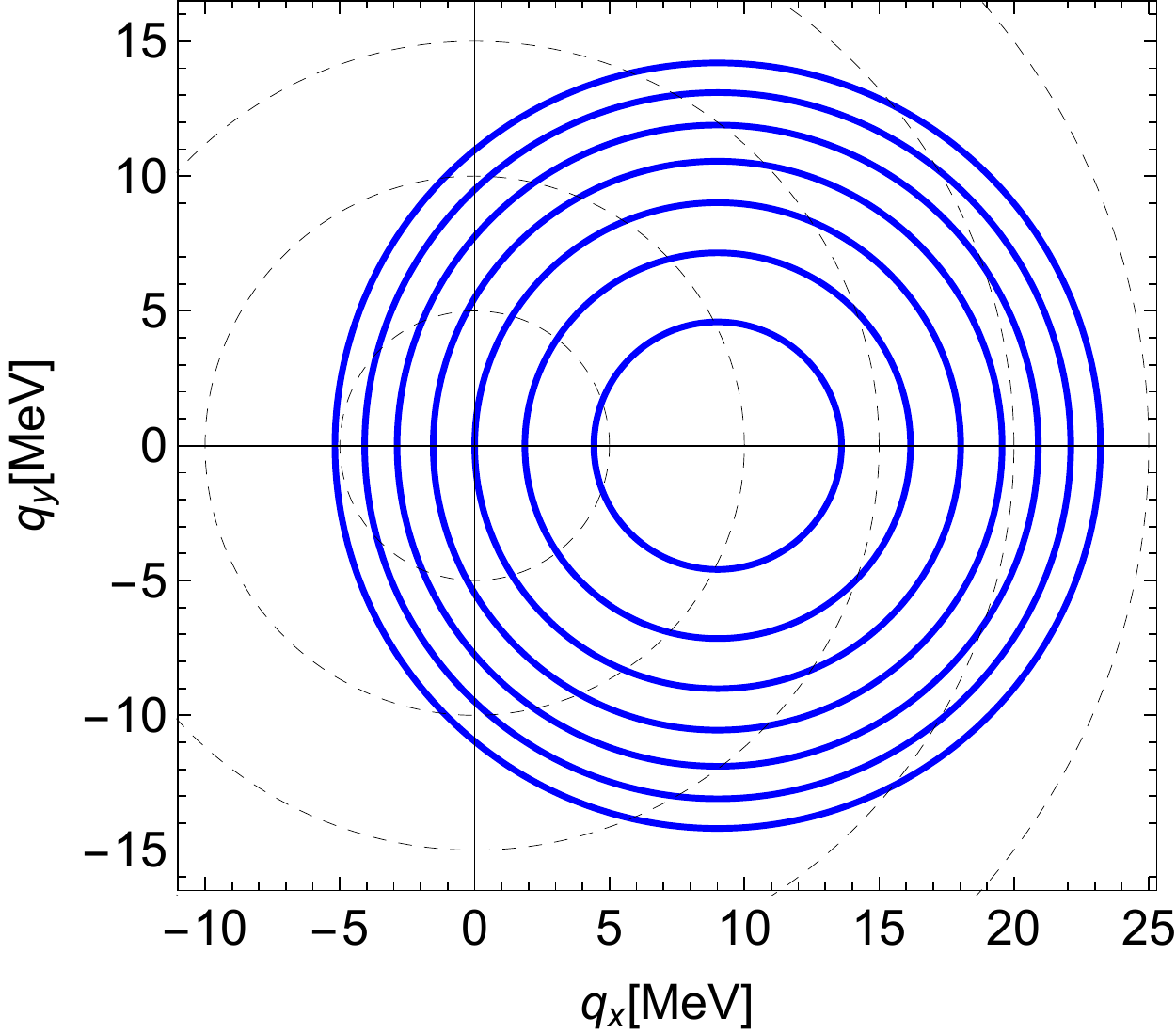}
\caption{Figure showing the origin of the ring in recoil momentum, $\vect{q}$, space for an F nucleus due to a collision with a $m_\chi = 100$~GeV WIMP with velocity $\vect{v}$. Here $q_x=q \cos{\theta}$ and $q_y=q \sin{\theta}$ are the components  of the recoil momentum $\vect{q}$ in the direction of the incoming WIMP velocity $\vect{v}$ and perpendicular to this direction respectively.}
\label{RingOrigin}
\end{figure}

The origin of the ring can be understood considering just one incoming WIMP with velocity $\vect{v}$. Energy and momentum conservation in the collision of this WIMP with a target nucleus of mass $m_T$  imply that the magnitude of the recoil nuclear momentum is  given by eq.~\eqref{eq:vq-inelastic} where $E_R= q^2/2 m_T$,  $v_q = v \cos{\theta}$ and $\theta$ is the scattering angle of the recoiling nucleus with respect to the incoming WIMP velocity. After these replacements, eq.~\eqref{eq:vq-inelastic}  becomes  eq.~\eqref{eq-q} (see appendix~\ref{app-A}). Solving  eq.~\eqref{eq-q} for $q$ we get
\beq
q =  \mu v \left[ \cos{\theta} \pm \sqrt{\cos^2{\theta} - \frac{2\delta}{\mu v^2} } \right].
\label{qextracted}
\eeq
As shown in figure~\ref{RingOrigin} for F recoils with $m_\chi = 100$~GeV WIMPS, in momentum space these values of $q$ lie on the surface of a sphere of radius $\mu v \sqrt{1 - (2\delta/\mu v^2)}$ centered at $\mu v$ (shown as blue circles in the figure for several values of $\delta$). For a certain recoil energy $E_R = q^2/2m_T$, $\vect{q}$ must be, in momentum space, on a sphere with radius $q = \sqrt{2 m_T E_R}$  centered at the origin $\vect{q }= 0$ (shown as dashed black circles in the figure for several values of $E_R$). The only possible $\vect{q}$ values at fixed magnitude $q$ are therefore on the ring on which the two spheres mentioned intersect. The angular radius of the ring, $\gamma$, fulfills the condition in eq.~\eqref{gamma}.

In the following we assume that the local DM velocity distribution is an isotropic Maxwell-Boltzmann distribution, truncated at $v_{\rm esc}$, with a velocity dispersion independent from the local circular speed. The range of the best fit peak speeds of the Maxwellian distribution from hydrodynamic simulations is 223 -- 289 km$/$s~\cite{Bozorgnia:2016ogo}. This translates into a 3D velocity dispersion of  273 -- 354 km$/$s. Since the best prospects to observe the ring is for small $\sigma_v$~\cite{Bozorgnia:2011vc}, here we take $\sigma_v=273$~km$/$s. Following ref.~\cite{Bozorgnia:2011vc}, for the local circular speed, $v_c$, we take 180 km/s and 312 km/s as low and high estimates. This results in the minimum and maximum values of $V_{\rm lab}$ of 179.8 km/s and 340.2 km/s, respectively (see table 1 of \cite{Bozorgnia:2011vc}). The Galactic escape speed at the Solar position found by the RAVE survey is $v_{\rm esc} = 533^{+54}_{-41}$ km/s at the 90\% confidence~\cite{Piffl:2013mla}. Thus we adopt the median value, $v_{\rm esc}=533$~km/s.

\begin{figure}
\includegraphics[width=0.5\textwidth]{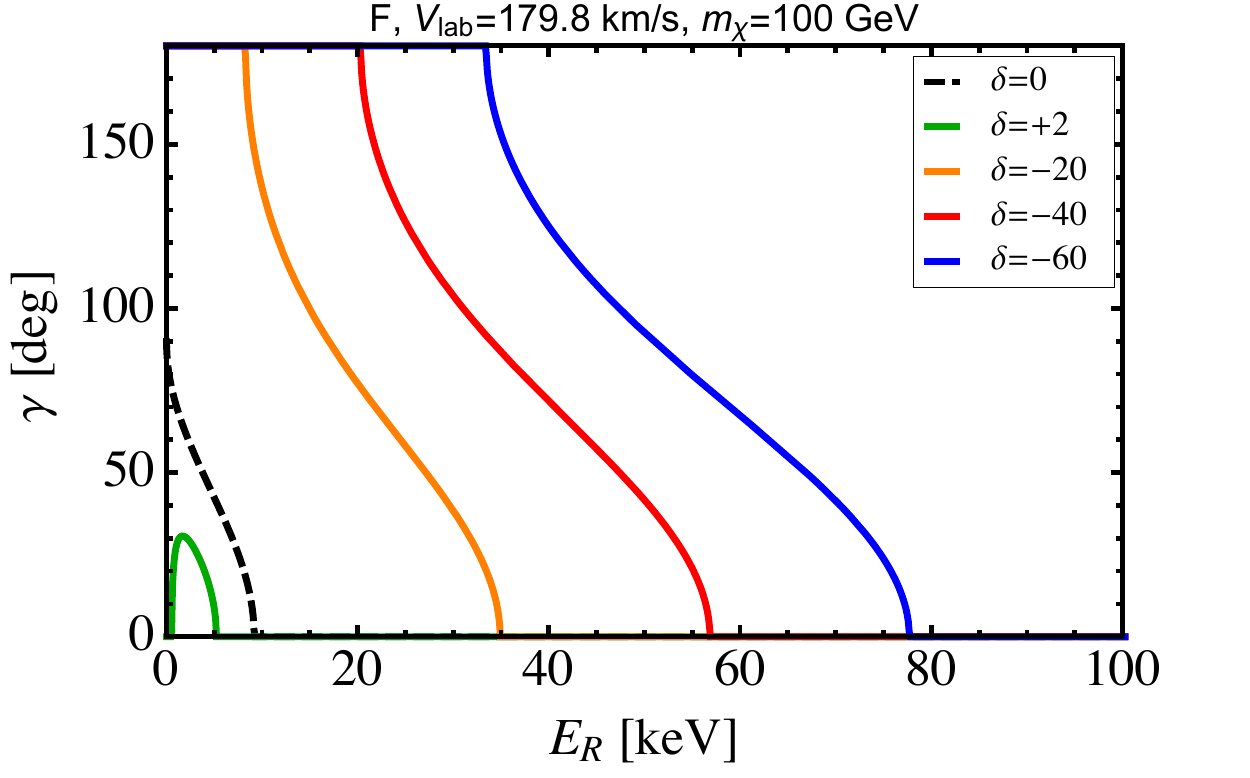}
\includegraphics[width=0.5\textwidth]{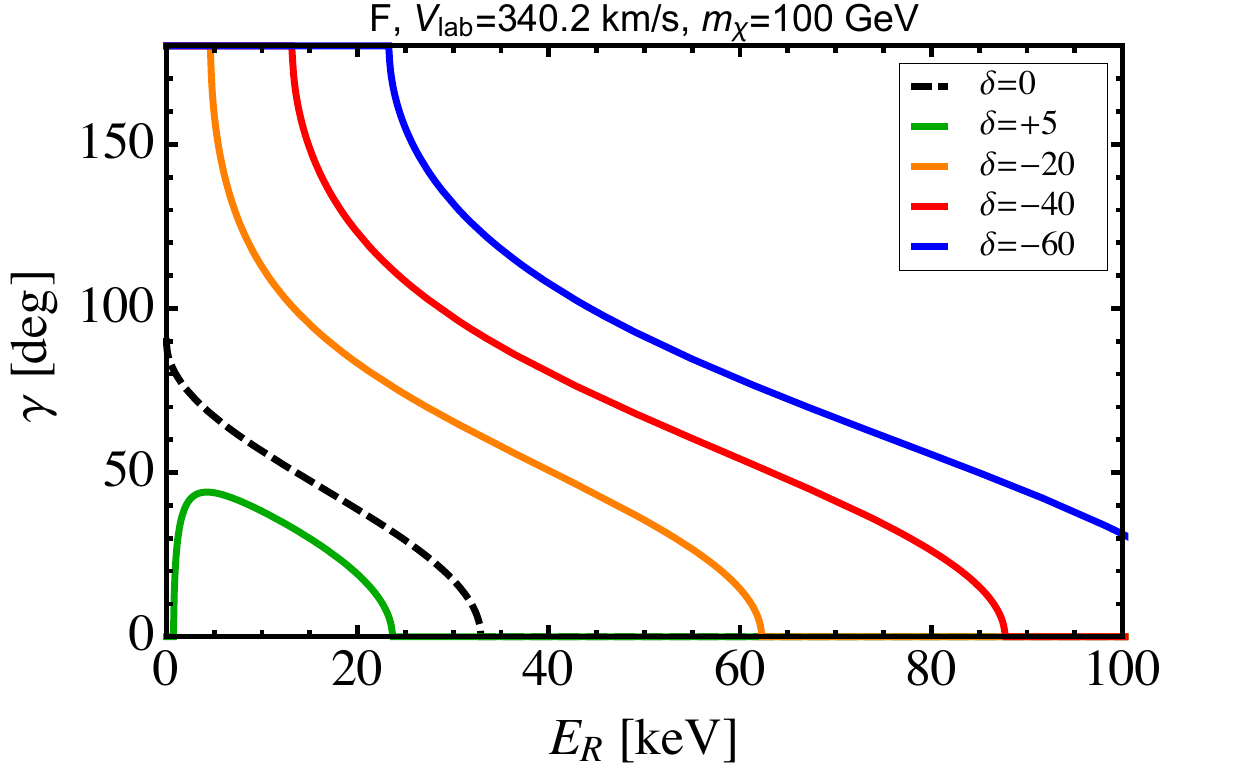}
\caption{Ring angular radius $\gamma$ as a function of $E_R$ for target element F for a 100 GeV WIMP mass. Left: $V_{\rm lab} =179.8$~km/s; right: $V_{\rm lab} =340.2$~km/s. The different color curves correspond to different values of the DM mass splitting $\delta$ in units of keV. Notice that for $\delta<0$, $\gamma=180^\circ$ at low recoil energies, and reaches zero at high recoil energies.}
\label{RingAngle}
\end{figure}

Figure~\ref{RingAngle} shows plots of $\gamma$ as a function of $E_R$ for various cases of positive and negative DM mass splitting $\delta$ for F recoils with $m_\chi = 100$~GeV and the low and high estimates of $V_{\rm lab}$. The black dashed line is for the case of elastic scattering. For exothermic scattering ($\delta < 0$), $\gamma=180^\circ$ at low recoil energies and the dipole feature is backwards. As the recoil energy increases, $\gamma$ decreases, and the ring-like feature appears in the backwards direction for $90^\circ<\gamma<180^\circ$. As the energy increases further, and for $0<\gamma<90^\circ$ the ring appears in the forward direction, similarly to the case of elastic scattering. Finally, for large enough recoil energies, $\gamma=0$, and the usual dipole feature in the forward direction appears. For endothermic scattering  ($\delta > 0$), $\gamma$ decreases to zero at low recoil energies, and a peak in $\gamma$ appears where there is the maximum ring radius. For $\delta > 2$~keV and $\delta>5$~keV in the left and right panels respectively, $v_q$ never becomes smaller than $V_{\rm lab}$ and the ring doesn't exist, i.e. $\gamma=0$.

\begin{figure}
\centering
\includegraphics[width=0.49\textwidth]{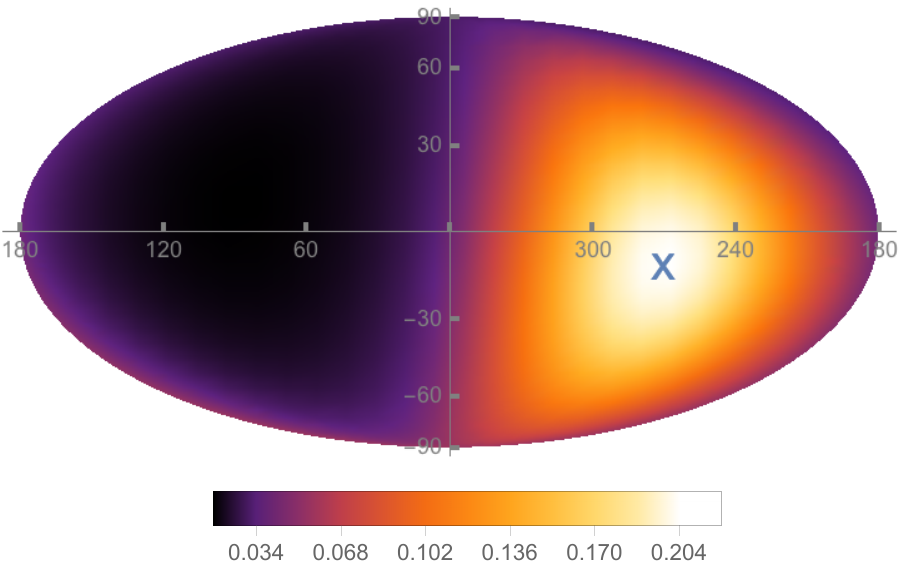}
\includegraphics[width=0.49\textwidth]{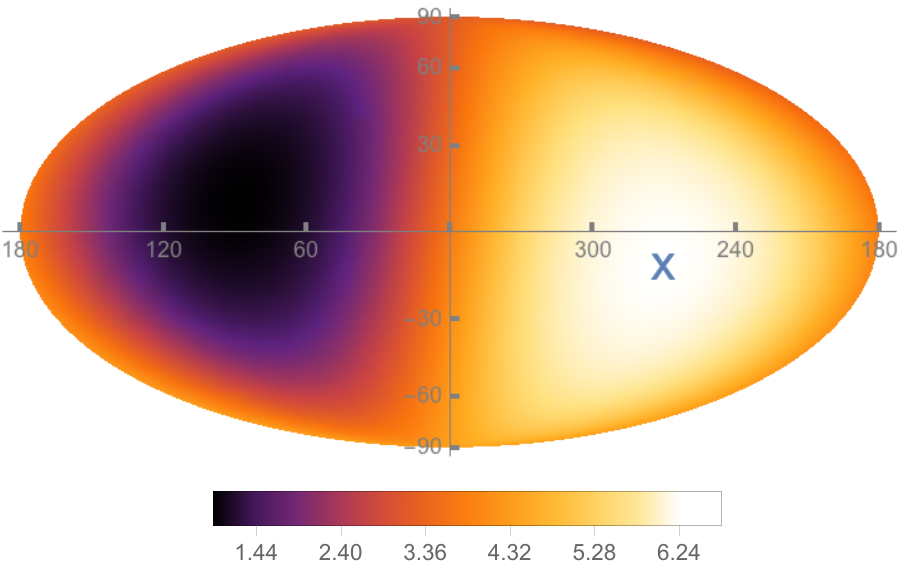}\\
\includegraphics[width=0.49\textwidth]{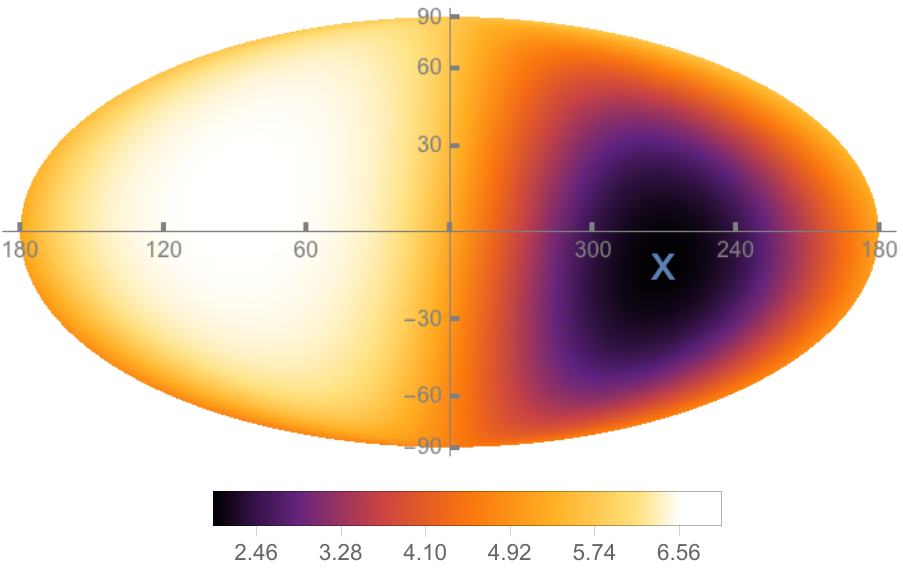}
\includegraphics[width=0.49\textwidth]{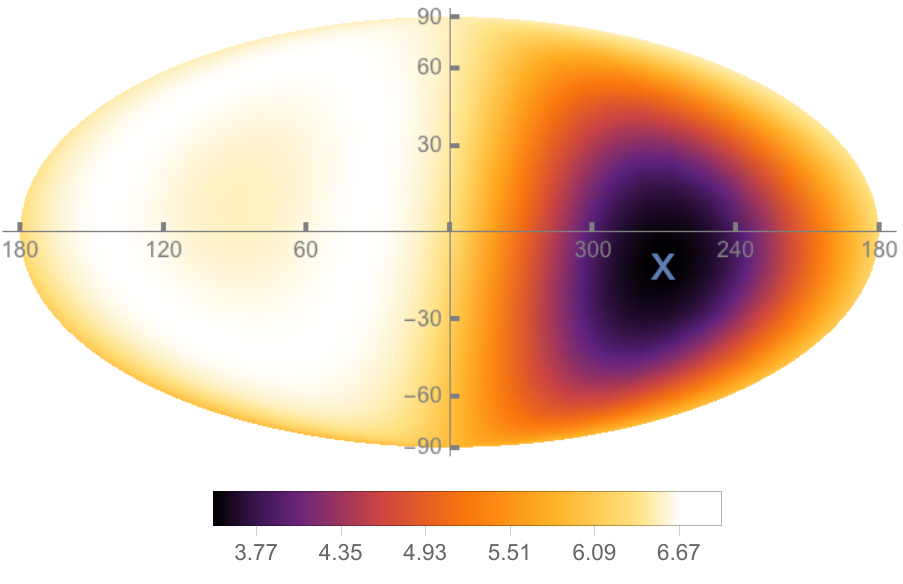}
\caption{Mollweide equal-area projection maps of the celestial sphere in Galactic coordinates showing (a) the number fraction $F_{\rm DM}(\hat{\vect v}, v_{\rm min})$ of 100 GeV WIMPs crossing the Earth per unit solid angle as a function of the WIMP velocity direction $\hat{\vect v}$. We assume $v_{\rm min}=185$ km/s, necessary to produce 20 keV fluorine recoils for $\delta=-40$ keV. The other panels show the directional differential recoil rate $dR/dE_Rd\Omega_q$ in F for (b) $\delta=0$ and $E_R=20$~keV, (c) $\delta=-40$~keV and $E_R=20$~keV, and (d) $\delta=-40$~keV and $E_R=25$~keV. In all panels we assume $v_c=180$~km$/$s, $\sigma_v=273$~km$/s$, $v_{\rm esc}=533$ km/s, and $m_\chi=100$~GeV on December 2. The direction
of $-\vect{V}_{\rm lab}$ is specified with  a cross on the maps. The values corresponding to each color shown in the horizontal bars are given in units of sr$^{-1}$ in panel (a), and in units of
$10^{-7} \times (\rho_{0.3} \sigma_{44}/{\rm kg~day~keV~sr})$ in the other panels, where $\rho_{0.3}$ is the DM density in units of 0.3 GeV cm$^{-3}$ and $\sigma_{44}$ is the WIMP-proton
cross section in units of $10^{-44}$~cm$^2$. In panel (d), the angular radius of the ring  is $\gamma = 126^\circ$.}
\label{fig:Mollweide}
\end{figure}

In figure~3 we illustrate the effect we point out in this paper. In figure~\ref{fig:Mollweide}.a  (top left panel) we show the arrival direction of 100 GeV WIMPs in the detector. In particular, we show the Mollweide equal-area projection map of the number fraction of WIMPs crossing the detector per unit solid angle with speed larger than $v_{\rm min}$, as a function of the WIMP velocity direction $\hat{\vect v}$,
\beq
F_{\rm DM}(\hat{\vect v}, v_{\rm min}) =  \int_{v_{\rm min}}^{v_{\rm max}(\hat{\vect v})} dv~v^2 f({\vect v}),
\label{qextracted}
\eeq
where $f({\vect v})$ is the DM velocity distribution in the detector reference frame, and $v_{\rm max} (\hat{\vect v})= - \hat{\vect v} \cdot {\vect V}_{\rm lab} + \sqrt{(\hat{\vect v} \cdot {\vect V}_{\rm lab} )^2 - {\vect V}_{\rm lab}^2 + v_{\rm esc}^2 }$.  Notice that the upper limit of the integral, $v_{\rm max} (\hat{\vect v})$, is such that $\left|{\vect v}+{\vect V}_{\rm lab}\right|=v_{\rm esc}$. The maximum of $F_{\rm DM}(\hat{\vect v}, v_{\rm min})$ occurs at $\hat{\vect v} \cdot {\vect V}_{\rm lab}=-V_{\rm lab}$, i.e.~in the direction of the average WIMP velocity $-{\vect V}_{\rm lab}$, marked with a cross in the figure. In figure~\ref{fig:Mollweide}.a we assume $v_{\rm min}=185$ km/s, necessary to produce 20 keV fluorine recoils for $\delta=-40$ keV.

In figures~\ref{fig:Mollweide}.b (top right panel), \ref{fig:Mollweide}.c (bottom left panel), and \ref{fig:Mollweide}.d (bottom right panel) we show Mollweide equal-area projection maps of the directional differential recoil rate in fluorine for different types of scattering with 100 GeV WIMPs.  Figure~\ref{fig:Mollweide}.b is for the case of elastic scattering, $\delta=0$, and the usual dipole feature in the forward direction, the average direction of WIMP arrival, is visible. In figures~\ref{fig:Mollweide}.c and \ref{fig:Mollweide}.d we assume $\delta=-40$ keV, and the maximum of the recoil rate appears in the backward direction. The backward dipole (ring-like) feature is visible in figure~\ref{fig:Mollweide}.c  (\ref{fig:Mollweide}.d) where we assumed a 20 keV  (25 keV) recoil energy. Notice that in figures~\ref{fig:Mollweide}.c and \ref{fig:Mollweide}.d the scattering is purely exothermic because the minimum required WIMP speed for endothermic scattering (with $\delta= 40$ keV) is $v_\delta=$ 692 km/s, larger than the assumed value of the escape speed $v_{\rm esc}=$ 533 km/s.

\section{Conclusions}
\label{Conclusions}

The observation of a preferred nuclear recoil direction with respect to the Galaxy in directional dark matter detection would be an unmistakable  signature of the dark matter. It is usually assumed that the average
recoil direction should coincide with the average WIMP arrival direction on Earth. Here we point out that this is not always true for inelastic exothermic dark matter. For this type of dark matter  the mean recoil direction as well as a secondary feature, a ring of  maximum recoil rate around the  mean recoil direction, could instead be opposite to the average dark matter  arrival direction. Thus, the detection of an average nuclear recoil direction opposite to the usually expected direction of WIMP arrival would constitute a spectacular experimental confirmation of this type of dark matter. 

\subsection*{Acknowledgements}

N.B.~acknowledges support from the European Research Council
through the ERC starting grant WIMPs Kairos.  G.G.~was supported in part by the  US Department of Energy Grant DE-SC0009937. P.G.~was supported by NSF award PHY-1415974. G.G.~thanks the hospitality of the Department of Physics of the University of Padova, Italy, where part of this work was done.

\appendix

\section{Radon transform for inelastic scattering}
\label{app-A}

In this appendix we derive the expression of $v_q$ entering the Radon transform for the case of inelastic scattering. We perform the full relativistic kinematics in the inelastic case (to be precise, in the case in which the outgoing particle has mass $m_{\chi^*}$ not necessarily equal to the mass $m_{\chi}$ of the incoming DM particle, and the nucleus recoils without excitation). We write down the conservation of energy and momentum relativistically.

In the lab frame (nucleus initially at rest) the quantity $\vect{p} \cdot \vect{q}$, where $\vect{p}$ is the DM incoming momentum and $\vect{q}$ is the outgoing nucleus momentum, is
\beq
\vect{p} \cdot \vect{q} = \frac{(m_{\chi^*}^2-m_{\chi}^2)}{2} + (E+m_T) E_R,
\eeq 
where $E = m_\chi (1-v^2)^{-1/2}$ is the initial energy of the incoming particle, and $E_R$ is the relativistic kinetic energy of the nuclear recoil, $E_R = \sqrt{m_T^2+q^2}-m_T$. In the non-relativistic case we are considering, 
\beq
E_R = \frac{q^2}{2m_T} .
\eeq

This gives for $v_q = \vect{v} \cdot \hat{\vect q}$, which is the quantity appearing in the Radon transform (here $\vect{v}$ is the incoming DM velocity),
\beq
v_q = \frac{(m_{\chi^*}^2- m_{\chi}^2)}{2Eq} + \left(\frac{E+m_T}{Em_T} \right) \left( \frac{m_TE_R}{q} \right).
\eeq
In the non-relativistic limit, $E \rightarrow m_{\chi}$ and $m_TE_R/q \rightarrow q/2$, so
\beq
v_q  = \frac{m_{\chi^*}^2- m_{\chi}^2}{2mq} + \frac{q}{2\mu}
\eeq
or
\beq
v \cos \theta  = v_q  = \frac{1}{q} \left[ \frac{q^2}{2\mu} + \delta + \frac{\delta^2}{2 m_{\chi}} \right].
\label{eq-q}
\eeq
Neglecting the $\delta^2$ term, we recover eq.~\eqref{eq:vq-inelastic}.

\bibliographystyle{JHEP.bst}
\bibliography{biblio}

\end{document}